\documentclass[11pt]{article}
\usepackage{fullpage}
\usepackage{graphicx}
\usepackage{hyperref}
\usepackage{color}

\usepackage{inputenc}
\usepackage{subcaption} % Provides subfigure; Requires label _after_ caption
\usepackage{listings}
\usepackage{url}
\usepackage{comment}
\usepackage{fancybox}
\usepackage{xcolor}
\usepackage{algorithm}
\usepackage{algpseudocode}
\algtext*{EndWhile}% Remove "end while" text
\algtext*{EndIf}% Remove "end if" text
\algtext*{EndFunction}% Remove "end function" text
\algnewcommand\True{\textbf{true}\space}
\usepackage{enumitem}  % Options of enumitem can make lists more compact
\newcommand{\fixme}[1]{\textcolor{red}{#1}}

\lstdefinestyle{code_style}{
    basicstyle=\ttfamily\footnotesize,
    breakatwhitespace=false,         
    breaklines=true,                 
    captionpos=b,                    
    keepspaces=true,                 
    numbers=left,                    
    numbersep=5pt,                  
    showspaces=false,                
    showstringspaces=false,
    showtabs=false,                  
    tabsize=2
}

\title{Enabling Practical Transparent Checkpointing for MPI: \\
       A Topological Sort Approach}
\author{
Yao Xu \\
Khoury College of Computer Sciences \\
Northeastern University \\
Boston, USA \\
xu.yao1@northeastern.edu
\and
Gene Cooperman \\
Khoury College of Computer Sciences \\
Northeastern University \\
Boston, USA \\
gene@ccs.neu.edu
}
\date{}

\begin{document}

\maketitle

\begin{abstract}
MPI is the de facto standard for parallel computing on a cluster of
computers. Checkpointing is an important component in any strategy for
software resilience and for long-running jobs that must be executed by
chaining together time-bounded resource allocations. 
This work solves
an old problem: a practical and general algorithm for transparent
checkpointing of MPI that is both efficient and compatible with most of
the latest network software.
Transparent checkpointing is attractive due to its generality and
ease of use for most MPI application developers.
Earlier efforts at transparent checkpointing for MPI, one decade ago,
had two difficult problems: (i)~by relying on a specific MPI
implementation tied to a specific network technology; and (ii)~by
failing to demonstrate sufficiently low runtime overhead.

Problem~(i) (network dependence) was already solved in 2019 by MANA's introduction
of split processes. Problem~(ii) (efficient runtime overhead) is solved
in this work.
This paper introduces an approach that avoids these limitations,
employing a novel topological sort to algorithmically determine a safe
future synchronization point.  The algorithm is valid for both blocking
and non-blocking collective communication in MPI.
We demonstrate the efficacy and scalability
of our approach through both micro-benchmarks and a set of five real-world MPI
applications, notably including the widely used VASP (Vienna Ab Initio
Simulation Package), which is responsible for 11\% of the workload on the
Perlmutter supercomputer at Lawrence Berkley National Laboratory. VASP was
previously cited as a special challenge for
checkpointing, in part due  to its multi-algorithm codes. 

\end{abstract}

\section{Introduction}
\label{sec:introduction}

MPI is a de facto standard for multi-node computing in modern HPC sites.
Checkpointing of MPI applications has become essential for long-running
computations, due to concerns for software resilience, and especially
for \emph{chaining of resource allocations}.  Because of reasons of system
maintenance (both scheduled and unscheduled), a single resource allocation
has a maximum time limit, such as 48~hours.

This work demonstrates a \emph{collective clock} (CC) algorithm for topological sorting. This represents the first algorithm that can efficiently and transparently checkpoint MPI applications under the newest network architectures.
The work relies on the split-process approach originally introduced with the original MANA prototype (MPI-Agnostic, Network-Agnostic checkpointing)~\cite{garg2019mana}.
This work demonstrates for the first time a novel algorithm that achieves low runtime overhead,
while supporting the newest network interconnects.
This is in contrast with MANA, which had a fatal flaw in terms of
runtime overhead, when applied to codes making intensive use of
MPI collective operations (e.g., \texttt{MPI\_Barrier}, \texttt{MPI\_Bcast}).

This work showcases the application VASP (Vienna Ab Initio Simulation
Package)~\cite{hafner2008ab}. On Perlmutter, the \#14 supercomputer
in the world~\cite{top5002024jun}, VASP is responsible for 11\% of the
CPU cycles~\cite{li2023analyzing}.

VASP was cited as a future challenge for low-overead transparent
checkpointing, in a 2021 study~\cite[Section~IV-B]{xu2021mana}.
As seen in Table~II of Section Section~IV-B of the study, the VASP workload CaPOH with 4 nodes (128 MPI processes) was shown to incur a runtime overhead of 40\% (35~seconds runtime with MANA versus 25~seconds for running VASP natively).
That study was run on the Cori supercomputer, a predecessor to the current Perlmutter at NERSC.

The importance of low-overhead transparent checkpointing for VASP
is explained by its heavy dependence on FFTs.
Efficient FFTs depend heavily on low-latency networks.
As VASP runs on more nodes (e.g., above 4~nodes),
the workload efficiency degrades, due to a greater fraction of
inter-process communication being across the network to a different node.
Hence, long-running VASP sessions prefer to run on a small number of nodes, and compensate for the fewer nodes by running for a long time,
using chaining of resource allocations through checkpoint-restart.
This reliance of VASP on FFRs was confirmed by randomly stopping the
computation under GDB, and observing the process stack.

To place transparent checkpointing in perspective, there exist two broad alternatives for checkpointing for MPI:
\begin{enumerate}
\item transparent checkpointing; and
\item application-level checkpointing.
\end{enumerate}

The alternative of application-level checkpointing~\cite{moody2010design,bautista2011fti,bland2013post,laguna2016evaluating,nicolae2019veloc}
tends to be used on large codes that are memory-bound (that use most of the RAM on a node).
In contrast, transparent checkpointing is attractive for the remaining CPU-bound MPI codes, due to its generality and ease-of-use.
Unlike the appliction-level case, in transparent checkpointing MPI application, the binary does not have to be modified,
re-compiled, or re-linked.

Nevertheless, transparent checkpointing for MPI faced two fundamental obstacles:
\begin{comment}
    \begin{description}[nosep]
  \item[i. (network dependence):] later MPI implementation used newer network interconnects; and
  \item[ii. (runtime overhead):] unacceptably high runtime overhead.
\end{description}
\end{comment}

\begin{enumerate}
    \item \textbf{network dependence:} later MPI implementation used newer network interconnects; and
    \item \textbf{runtime overhead:} unacceptably high runtime overhead.
\end{enumerate}

Historically, a decade was spent unsuccessfully attempting production-quality, transparent checkpointing in the
2000s and early 2010s~\cite{hargrove2006berkeley,gao2006application,hursey2009interconnect,cao2014transparent,cao2016system}.
Those activities were primarily targeted toward the dominant network at that time:  OFED InfiniBand.  Those attempts eventually foundered on the proliferation of newer HPC networks: HDR Mellanox InfiniBand, Cray GNI, Intel Omni-Path,
HPE Slingshot-10, and HPE Slingshot-11.  In these cases, the earlier approach was abandoned in the face of HPC sites upgrading to newer network interconnects.

After an intervening decade, the first fundamental obstacle (\emph{network dependence}) was solved by the \emph{split process} technique, introduced by MANA~\cite{garg2019mana}.
The MPI application and the
underlying MPI library (and associated network library) were considered as independent
programs.  They were loaded independently into a single process.  The MPI application was compiled with MPI wrapper functions, which were bound \emph{at runtime} to the MPI library of the second program.  At checkpoint time, only the MPI application program was checkpointed.  At restart time, the second program (MPI library and network) was loaded, and then at runtime, it determined its own MPI rank in the world communicator, and then loaded the checkpointing image of the MPI application for the given rank.

As discussed in~\cite{xu2021mana}, while the transparent checkpointing approach had successfully been brought back to life, there remained the fundamental problem of high runtime overhead.  The second problem appeared primarily in MPI codes that intensively called MPI collective operations.

In particular, the high runtime overhead of VASP remained a challenging problem for checkpointing.  This heavy reliance of VASP on MPI collective operations is seen in Figure~4 of~\cite[Section~IV-B]{xu2021mana}:  The number of collective MPI operations per second rises drastically, from 4,800 collective operations per second for 128~MPI processes to 7,000 collective operations per second for 256~MPI processes).

This work solves the second fundamental problem, \emph{high runtime overhead}.
The novel \emph{collective clock} (CC) algorithm uses multiple sequence numbers
to track, for each MPI group, how many collective calls have been iterated so far.
Incrementing a sequence number for the MPI group of an MPI collective operation does not require network communication, and is inherently low overhead (see the micro-benchmarks in Figures~\ref{fig:micro_benchmark} and~\ref{fig:overlaps}).
In the micro-benchmarks and five real-world applications, the typical runtime overhead is from 0\% to 5\%.

This enhancement of MANA is open source.  It can currently be found at~\cite{mana2024dynamic}, and will soon replace the current main branch of MANA~\cite{mana2024main}.  The core of the CC algorithm in the new branch can be found in the file \texttt{seq\_num.cpp}.

Reducing runtime overhead for transparent checkpointing has become especially important in the most recent generation of very high performance network interconnects.  For example, a modern interconnect like HPE Slingshot-11 on the Perlmutter supercomputer enables up to a quarter of a million collective calls per second (see Table~\ref{tbl:inputs}; OSU micro-benchmark running on four computer nodes).  Earlier supercomputer (and therefore earlier MPI codes) used earlier network interconnects such as Ethernet and OFED InfiniBand, and so did not experience such high rates of collective calls.  Therefore, since communication efficiency was not under stress, earlier MPI codes tended to prefer the simpler point-to-point communication calls of MPI, and made only limited use of collective communication.

Newer or revised MPI applications are intended to run on newer network interconnects, and therefore tend to require the more efficient CC algorithm for reduced runtime overhead.
Table~\ref{tbl:inputs} of Section~\ref{sec:experiments} shows the rate of collective communication calls per second across a variety of applications.

\subsection{Points of Novelty}

This work introduces a new collective-clock algorithm for checkpointing MPI.
The central points of novelty are:
\begin{enumerate}
    \item a novel topoligical sort algorithm for collective communication;
    \item the first transparent checkpointing algorithm that
      supports MPI's \emph{non-blocking} collective operations,
      needed for overlapping computation and communication (\hbox{Section}~\ref{sec:nonblocking}); and
    \item an implementation of transparent checkpointing for MPI that avoids the
    high runtime overhead of Garg \hbox{et al.}~\cite{garg2019mana},
    which had inserted an additional \texttt{MPI\_Barrier}
    in front of each collective operation.
\end{enumerate}

While this work is intended primarily for MPI in HPC, the lessons may also apply
to algorithms in distributed computing.  For example, the MPI 
collective routines \texttt{MPI\_Reduce} and \texttt{MPI\_Scan} are 
closely related to parallel prefix~\cite{blelloch1989scans},
and there may be extensions to concepts of~\emph{atomic broadcast}~\cite{cristian1995atomic}.

\subsection{Organization of Paper}

The organization of this work is as follows.
Section~\ref{sec:background} provides brief background on MPI itself and the split-process architecture that forms the basis for MANA to checkpoint MPI.
Section~\ref{sec:mpi-standard} reviews some essential points in the semantics of MPI.
Section~\ref{sec:seq_num} presents a novel collective-clock (CC) algorithm, which replaces MANA's original two-phase-commit algorithm.
Section~\ref{sec:experiments} provides an experimental evaluation.
Section~\ref{sec:relatedWork} discusses the related work, and
Section~\ref{sec:conclusion} presents a conclusion.

\section{Background}
\label{sec:background}

Section~\ref{sec:mpi} reviews basic concepts for MPI itself, while Section~\ref{sec:2pcSplitProcess} reviews the \emph{split process} mechanism that is the basis for the MANA architecture.  The CC algorithm described here replaces the original two-phase-commit algorithm that was introduced in the original MANA paper~\cite{garg2019mana}.

\subsection{Review of MPI}
\label{sec:mpi}
Each MPI process has a unique rank as an MPI-specific process id in a given communicator.  MPI provides point-to-point operations on the ranks, such as \texttt{MPI\_Send} and \texttt{MPI\_Recv}. MPI also provides collective operations.  Examples include \texttt{MPI\_Barrier}, \texttt{MPI\_Bcast}, \texttt{MPI\_Alltoall}, etc.  A \emph{collective MPI operation} is executed in parallel by a subset of the MPI processes.  In this case, each member of the subset of processes for that operation must individually make a corresponding MPI call, to successfully invoke the parallel operation.

The subset of MPI processes participating in an operation is referred to as an \emph{MPI group}.  An\linebreak[4]
\emph{MPI\_Communicator} can be created from an MPI group.  Creating a communicator is a parallel operation in which each participating process receives a handle to a common communicator representation, shared by all participating processes.
For a single MPI collective operation, all participating processes must call the same MPI collective function with the same MPI communicator.

An initial communicator \texttt{MPI\_COMM\_WORLD} includes all the MPI processes.  Each new group numbers its processes consecutively, beginning with rank~0.
\texttt{MPI\_Group\_translate\_ranks} is available to determine the rank of a process within a new group, as compared to a previous group.

Finally, MPI also defines non-blocking variants, such as \texttt{MPI\_Isend}, \texttt{MPI\_Irecv}, \texttt{MPI\_Ibcast}, \texttt{MPI\_Ibarrier} and \texttt{MPI\_Ialltoall}.
 The variants have an additional argument, a pointer to an MPI \emph{request} object.
and does not block.

A non-blocking call first \emph{initiates} the MPI operation. Then non-blocking call immediately sets the request object and returns. MPI may or may not
immediately begin executing the operation, depending on the implementation.
This enables overlap of communication and computation.

An individual MPI process tests if its part in the operation is \emph{locally complete} by calling \texttt{MPI\_Test}, \texttt{MPI\_Wait}, or a related call. 
The calls to test for completion include both the MPI request and a completion flag argument.
The current process has completed its part in the operation if the flag is set to true.
The request object is then modified in place to set its value to the pseudo-request \texttt{MPI\_REQUEST\_NULL}.

\subsection{MANA's Split Process Software Architecture and Checkpointing}
\label{sec:2pcSplitProcess}

\begin{figure}[h!t]
\centering
\includegraphics[width=0.7\columnwidth]{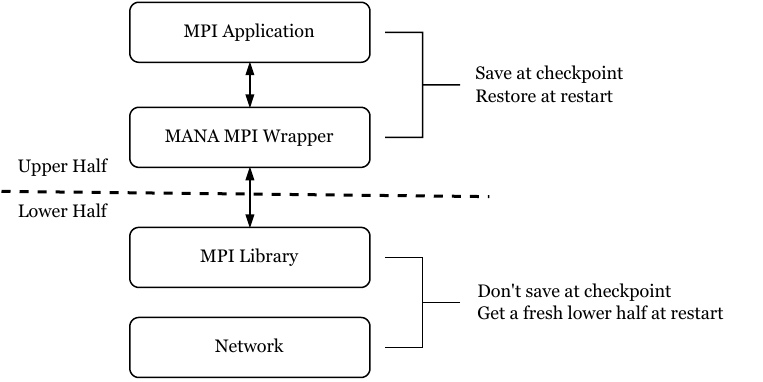}
\caption{Split Process Achitecture}
\label{fig:split_process}
\end{figure}

The current work adopts MANA's \emph{split process architecture}, which is illustrated in Figure~\ref{fig:split_process}. The split process architecture was introduced with
MANA (MPI-Agnostic Network-Agnostic transparent checkpointing)~\cite{garg2019mana}.
Two programs are loaded into a single memory space. The upper half contains the
MPI application and a library of wrapper ``stub'' functions that redirect MPI calls to the lower half. The lower
half contains a proxy program that is linked with network and MPI libraries. On restart the
upper half is restored and the lower half is replaced by a new one. This design
decouples the MPI application from the underlying libraries that ``talk'' to the hardware.

When MANA takes a checkpoint, it saves only the memory regions associated with the
upper half.  When MANA restarts, it begins a new ``trivial'' MPI application with
the correct number of MPI processes.  Each MPI process becomes the new lower half,
and it restores to memory the upper-half checkpoint image file whose world rank is the same
as the world rank in the lower half.

Since MANA does not save the lower-half memory, it works independently of the
particular network interface, and independently of the MPI implementation (providing
that the MPI implementation obeys MPI's standard API).

Thus, a ``safe'' state for MANA to checkpoint must obey the following invariant:
\begin{quote}
    \textbf{Collective Invariant:} \textit{No checkpoint may take place while an MPI process is inside a collective communication routine in the lower half.}
\end{quote}

The original MANA paper proposed a two-phase-commit (2PC) algorithm to find a safe state
obeying the collective invariant above.
The core idea of the earlier two-phase-commit algorithm is to use a wrapper function
around each MPI collective call to insert a call to \texttt{MPI\_Barrier} (or a call
to \texttt{MPI\_Ibarrier} followed by a loop of calls to \texttt{MPI\_Test}).
    When it is
time to checkpoint, if all processes have entered the barrier, then MANA waits
until all processes have completed the collective call.  If some processes have
not yet entered the barrier, then it is safe to checkpoint because other processes 
cannot skip the barrier and start the real collective communication. 
On restart, if one had entered the barrier loop prior to checkpoint, then one calls 
\texttt{MPI\_Ibarrier} again before continuing. 

The severest limitation of the old 2PC approach is the high runtime overhead caused by the inserted barrier. The barrier forces all participating processes to synchronize. The synchronization takes extra time and blocks operations that are not required to be synchronized, such as \texttt{MPI\_Bcast}. In addition, forced synchronization conflicts with MPI's non-blocking collective communication model. Thus, non-blocking collective communications were not supported in the original MANA prototype.

\section{A Close Look at the MPI Standard}
\label{sec:mpi-standard}
A prerequisite to understand why the CC algorithm is correct requires
a precise reading of the MPI standard and its implications.
This document primarily cites the MPI-4.0
standard~\cite{message2021mpi}. 

Two key points from the MPI standard will
be used frequently in this work.
\begin{quotation}
\noindent
First, a correct, portable MPI program must assume that MPI collective
operations (blocking and non-blocking) are synchronizing.
Any program violating this assumption is erroneous.
\end{quotation}

The reason is that for any given MPI collective operation, a particular
MPI library may implement the collective operation as synchronizing.
Recall that \emph{synchronizing}~\cite[Section~3.2.12]{bangalore2019exposition}
means it acts as a barrier.
In a synchronizing call, no process can exit until all processes have
entered the call.
Hence, portable user programs must assume this more restrictive case.
Quoting the standard,
``a correct, portable program must invoke collective communications so that deadlock will not
occur, whether collective communications are synchronizing or not.''~\cite[Section~6.14]{message2021mpi}.

\begin{quotation}
\noindent
Second, once all participating MPI processes have initiated a non-blocking
operation, then the operation continues ``in background'', and must eventually
complete, independently of other actions by any of the MPI processes.
\end{quotation}

In the words of the standard,
``The progress of multiple outstanding non-blocking collective
operations is completely independent.''~\cite[Example~6.36]{message2021mpi}

\section{Collective Clock (CC) Algorithm}
\label{sec:seq_num}

The new algorithm is called CC due to its use of a happens-before relation on a vector of ``timestamps''.  In this sense, it bears a resemblance to the idea of logical clocks~\cite{baldoni2002fundamentals}.  The initial point of departure is that instead of employing logical clocks based on MPI processes, the \emph{collective clock} is based on logical clocks based on MPI communicators (in fact, on the underlying MPI groups).

The CC algorithm introduces a \emph{sequence number} for each group of MPI processes.  The sequence number is initialized to zero.  When a collective operation occurs on a group, the sequence number for that group is incremented locally.  When a checkpoint (the analog of a distributed snapshot) is requested, a \emph{target (sequence) number} is computed for each MPI group, as the maximum of the sequence number of that MPI group that is seen at each MPI process.  For the purposes of target
numbers, two MPI groups are considered to be the same if they satisfy \texttt{MPI\_SIMILAR}, meaning
that they contain the same set of MPI processes.
If an MPI process has never participated in that group for collective communication, then its sequence number is zero.

At checkpoint time, each MPI process constructs a set of target numbers for each MPI group in which the process participates.  The process then continues to execute until, for each MPI group containing that process, the process has reached the target number of that MPI group.

\begin{comment}
Note that a consistent CC snapshot requires the following property:
\begin{quotation}
\vskip-4ex
\begin{propA}
\hypertarget{prop:globalCollectives}{A collective operation} on an MPI group will cause the CC algorithm to assign the same sequence number to the group, locally, within each member process of the group. 
\end{propA}
\end{quotation}
Note that the MPI standard states:
``Collective operations can (but are not required to) complete as soon as the caller's participation in the collective communication is finished.''~\cite[Section~6.1]{message2021mpi}.
Hence, a particular MPI implementation is allowed to implement an operation as synchronizing.  So, an MPI program that deadlocks for synchronizing collective operations is non-portable, and therefore erroneous.  For further insights, see Example~6.27 of Section~6.14 of the MPI standard~\cite{message2021mpi}.
 
\end{comment}

\subsection{Definitions}
\label{subsec:definitions}
Some formal terms are defined next, and used in the description of the algorithm.
\begin{description}
\item[\textbf{Global group id (ggid):}] For the underlying MPI group of a communicator,
we assign a \emph{global group id (ggid)} based on hashing the ``world rank'' of each participating
MPI process according to its rank in MPI\_COMM\_WORLD. Communicator IDs generated by
the MPI library are local resource handles.  So we need to compute the ggid to identify communicators
globally. By design, similar communicators in the sense of \texttt{MPI\_SIMILAR} have the same ggid.
\begin{comment}
\\
\emph{(NOTE: The computation of the world rank of a participating process is made
efficient by applying the MPI routine
\emph{\texttt{MPI\_Group\_translate\_ranks}} (a local MPI operation)
to the local rank and MPI group for a given communicator.)}
\end{comment}

\item[\textbf{Sequence number ({SEQ[ggid]}):}] The (local) sequence number of a ggid (often
denoted \emph{SEQ[ggid]})
is a local, per-process counter that
records the number of calls to (blocking) collective communication routines using
that MPI group.  If a particular MPI process has never invoked a collective communication using the given
MPI group, then the local sequence number for that ggid is zero.
\item[\textbf{Target number ({TARGET[ggid]}):}] The  (global) target number of a ggid
(often denoted\linebreak[4] \hbox{\emph{TARGET[ggid]}})
is a global value, representing the maximum sequence number of the given ggid across each MPI process.
\item[\textbf{Reached a target}] A target is reached after we execute a blocking
collective call whose sequence number is equal to the target number.
\item[\textbf{Safe state}] A safe state of an MPI program in MANA is a state of execution
for which it's safe to checkpoint.
Two invariants must hold, to be in a safe state:

\textbf{Invariant 1:} \textit{No checkpoint must take place while a rank is inside a collective communication routine.}
This is the same as the collective invariant from Section~\ref{sec:2pcSplitProcess}.

\textbf{Invariant 2:} \textit{If a collective communication call has started when the checkpoint request
arrives, then the checkpoint request must be deferred until
all members of the communication can complete the
communication.}
\hbox{\ \ \ \ }
For example, suppose an \texttt{MPI\_Bcast} has started, and the sender process
has already broadcast its message.  Then the checkpoint
must be deferred for all receiving processes until they all can complete the communication.
\end{description}

\subsection{The CC Algorithm for Blocking Collective calls}
\label{sec:CCGeneral}

This subsection describes the CC algorithm in the context of blocking collective calls.  Later the interaction with blocking point-to-point calls will also be discussed.

Unlike the 2PC algorithm of MANA~\cite{garg2019mana}, the runtime overhead for CC remains almost zero
(see the micro-benchmarks of Section~\ref{sec:experiments}), since the only overhead is due
to interposing on MPI calls and incrementing
a sequence number.

\vskip25pt
\subsubsection{CC algorithm at runtime}\hfill

MANA provides a wrapper function for each collective communication call.
When a communicator is created, if the ggid of the underlying group has
not yet been seen, then the sequence number of that ggid is initialized:
\texttt{SEQ[ggid]=0}.  (See Section~\ref{subsec:definitions} for the definition
of the \emph{global group id} (ggid).)
During normal execution of an MPI application, each time a MANA wrapper
function is called on a blocking collective call, the global variable \texttt{SEQ[ggid]} for
that communicator is incremented.  No network operations are executed, and so
the runtime overhead in the CC algorithm remains exceptionally small.

\vskip25pt
\subsubsection{CC algorithm at checkpoint time}\hfill
At the time of checkpoint, the intuition behind CC is that execution should continue
until all MPI processes have reached a safe state.  A safe state is defined as
a state that satisfies the two safe-state invariants of Section ~\ref{subsec:definitions}.

The execution of an MPI computation can be viewed as a directed graph.  Each node corresponds to a collective communication call.  Each edge of the directed graph is labelled by an MPI process.  Each incoming edge of the node corresponds to the MPI process given by the edge label, entering the collective call.  Each outgoing edge corresponds to an MPI process exiting the collective call, and the edge label corresponds to that process.

The goal of the CC algorithm is to continue execution until:
\begin{enumerate}
    \item each node (collective call) that has been visited at checkpoint time by at least one MPI process will have been visited by all participating processes; and
    \item no other nodes have been visited.
\end{enumerate}

We define a node~$B$ to be \emph{dependent} on a node~$A$ if there is a directed edge from $A$ to~$B$.  The dependency relation is then extended by transitivity.
The extended directed graph is the directed graph that results when for each visited node, all participating processes have reached that node.
By transitivity, we define a node~$B$ to be \emph{dependent} on a node~$A$ if there is a path in the extended directed graph from~$A$ to~$B$.

The CC algorithm is a variant of a topological sort.  A topological sort dictates the order in which one visits a graph, such that for a node~$X$, all nodes that~$X$ is dependent on are visited before the node~$X$ is visited.  This is a restatement of condition~1, above.  In our setting, this is also the problem of finding a safe state for checkpointing.

More formally, we can state the following condition.  The process must be allowed to continue to execute if and only if:
\begin{quote}
    \emph{CONDITON $A$: the process has visited node~$A$, and node~$B$ is dependent on node~$A$, and some other process has already visited node~$B$.\hfill\hbox{\ }}
\end{quote}

The above condition~$A$ can be viewed as a condition on the sequence numbers that were described earlier.  When viewed in this way, Condition~A translates to saying the following.

For a given MPI process, the process should be allowed to proceed if and only if:
\begin{quote}
    \emph{CONDITON $A'$: the process is a member of a group whose global group id ($ggid$) is $a$ and $SEQ[a]<TARGET[a]$.\hfill\hbox{\ }}
\end{quote}

The interpretation is that if there's a node~$N$ in which $SEQ[a]<TARGET[a]$, then there will be a future node $M$ in which $TARGET[a]=SEQ[a]$.
Hence, the last visited node,~$N$, of this process is dependent on a future node~$M$ in the extended directed graph.

\begin{figure}[htb!]
\centering
\begin{subfigure}{0.31\linewidth}
\includegraphics[width=0.85\textwidth]{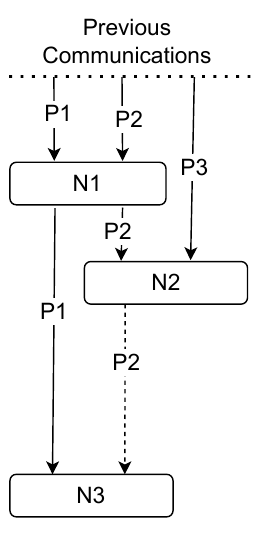}
\caption{Simple example of CC algorithm}
\label{fig:cc_directed_graph_simple}
\end{subfigure}
\hspace*{0.1\linewidth}
\begin{subfigure}{0.51\linewidth}
\includegraphics[width=0.9\textwidth]{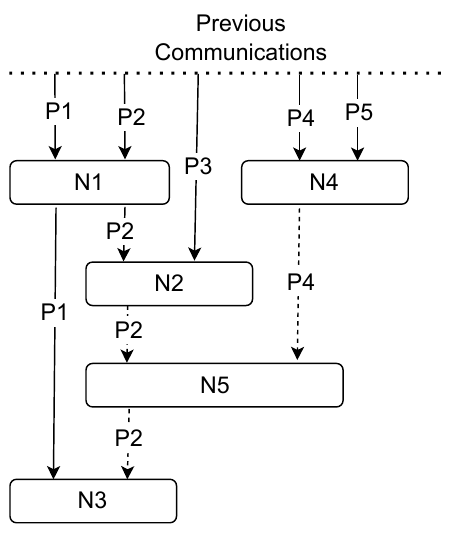}
\caption{example of CC algorithm with extended directed graph}
\label{fig:cc_directed_graph_complex}
\end{subfigure}
\caption{\small{Examples of the CC algorithm at checkpoint time. The execution of MPI communications is viewed as a directed graph. Each node corresponds to a collective communication. Each edge is labeled by an MPI process participating in the communication. Solid incoming edges indicate processes that already visited the node, whereas dotted incoming edges indicate future executions.  In Figure~(a), Condition~$A$ is applied once for $P2$ to continue executing.  In Figure~(b), $P2$ discovers the intermediate node~$N5$, and so Condition~$A$ is applied twice for $P2$ and once for $P4$.}}
\label{fig:cc_directed_graph}
\end{figure}

Figure~\ref{fig:cc_directed_graph} shows two examples of the CC algorithm at checkpoint time as 
directed graphs. Figure~\ref{fig:cc_directed_graph_simple} shows a simple case. At the time of
checkpoint, process 2 ($P2$) has visited $N2$. $N3$ is dependent on $N2$, and $N3$ has already been
visited by another process $P1$. According to \emph{Condition A}, $P2$ should continue executing until $P2$ also visits $N3$ to reach the safe state for checkpointing.

Figure~\ref{fig:cc_directed_graph_complex} shows a more complex case. Similar to the previous
example~\ref{fig:cc_directed_graph_simple}, $P2$ should continue executing to visit $N3$. In the process,
$P2$ needs to visit a new node $N5$, because $N3$ is also dependent on $N5$. On the other hand, $N5$ is dependent 
on another node $N4$. By applying \emph{Condition A} again, $P4$ should continue executing to visit $N5$.
Then, $P2$ is able to exit $N5$ and visit $N3$.

\subsubsection{The CC pseudo-code (blocking collective calls)}\hfill
\label{sec:pseudo-code}

It remains to show pseudo-code for the MPI function wrappers
used to implement the CC algorithm.  In Algorithm~\ref{algo:CCinitial}, each MPI
process exchanges sequence numbers ($SEQ[]$) and computes the targets ($TARGET[]$)
by calculating the global maximum sequence number for each $ggid$.

\setlength{\textfloatsep}{8pt}% Remove \textfloatsep
\setlength{\intextsep}{8pt}% Remove \textfloatsep

\begin{algorithm}[H]
\caption{Initial checkpoint request:  Initialize sequence number targets}
\begin{algorithmic}
\Function{Initialize\_seq\_num\_targets}{$\ldots$, MPI\_Comm $comm$}
\State $ckpt\_pending \gets$ \True
\ForAll {$G\in$ local MPI groups}
\State $P \gets$ MPI process members of $G$
\State $TARGET[ggid] \gets \max_P(SEQ[ggid])$
\EndFor
\EndFunction
\end{algorithmic}
\label{algo:CCinitial}
\end{algorithm}

\begin{comment}
\begin{algorithm}[htb!]
\caption{MANA's collective communication wrapper}
\begin{algorithmic}
\Function{Wrapper}{$\ldots$, MPI\_Comm $comm$}
\State \Call{commit\_begin}{$comm$}
\State Call the real MPI function
\State \Call{commit\_finish}{$comm$}
\State \Return Return value of the real MPI function
\EndFunction
\end{algorithmic}
\label{algo:wrapper}
\end{algorithm}
\end{comment}

\begin{algorithm}[htb]
\caption{After checkpoint request:  Interposing on collective communication via wrapper functions}
\begin{algorithmic}
\Function{Wrapper}{$\ldots$, MPI\_Comm $comm$}
\State \Call{Wait\_for\_new\_targets}{ } // Algorithm~\ref{algo:wait}
\State $ggid \gets$ translate // uses local fnc.: MPI\_Group\_translate()
\State Increment sequence number $SEQ[ggid]$
\If{$ckpt\_pending$ {\bf and} $SEQ[ggid]$>$TARGET[ggid]$}
  \State $TARGET[ggid] \gets SEQ[ggid]$
  \State \textbf{SEND:} Send $TARGET[ggid]$ to other members of $ggid$ group via MPI\_Isend
\EndIf
\State \textbf{EXECUTE:} Do collective communication call for this
                                                   wrapper function
\State \Call{Wait\_for\_new\_targets}{ } // Algorithm~\ref{algo:wait}
\EndFunction
\end{algorithmic}
\label{algo:CC}
\end{algorithm}

\begin{algorithm}[ht!]
\caption{Check if $SEQ[ggid]<TARGET[ggid]$ for some $ggid$, or if pending target updates}
\begin{algorithmic}
\Function{Wait\_for\_new\_targets}{\hbox{\ }}
\If{$SEQ[ggid]<TARGET[ggid]$ for some $ggid$}{\hbox{\ }\Return}
\EndIf
\While{$ckpt\_pending$}
\State $flag \gets$ MPI\_Iprobe(MPI\_ANY\_SOURCE,
                                  mana\_updates\_tag, mana\_comm, ...)
\If{$flag =$ \True}
\State $ggid \gets$ MPI\_Recv(..., mana\_updates\_tag, mana\_comm, ...)
\State \textbf{RECEIVE:} $TARGET[ggid] \gets$ MPI\_Recv(...,
                                    mana\_updates\_tag, mana\_comm, ...)
\If{$SEQ[ggid]<TARGET[ggid]$}{ \Return from function}
\EndIf
\EndIf
\EndWhile
\EndFunction
\end{algorithmic}
\label{algo:wait}
\end{algorithm}

Algorithm~\ref{algo:CC} is the implementation of Condition~$A'$ above.  Each MPI process executes until it has reached its target for all $ggid$.  While executing, an MPI process may increment $SEQ[ggid]$ for some $ggid$, such that $SEQ[ggid]>TARGET[ggid]$.  In that case, $TARGET[ggid]$ is updated, and the line highlighted by \textbf{SEND} in boldface then sends the new target to all members of the $ggid$ group.  The helper function, \texttt{Wait\_for\_new\_targets()} (Algorithm~\ref{algo:wait}) is called at the beginning and end of the wrapper function (Algorithm~\ref{algo:CC}), in order to receive and update any newly incremented targets.

\vskip25pt
\subsubsection{CC at checkpoint time:  An example}\hfill

\begin{figure}[htb!]
\centering
\begin{subfigure}{0.41\linewidth}
\includegraphics[width=\textwidth]{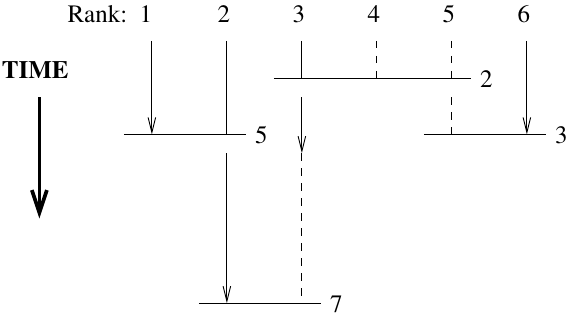}
\caption{Simple example of CC algorithm: setting the target sequence numbers}
\label{fig:seq-num-algo-simple}
\end{subfigure}
\hspace*{0.1\linewidth}
\begin{subfigure}{0.41\linewidth}
\includegraphics[width=\textwidth]{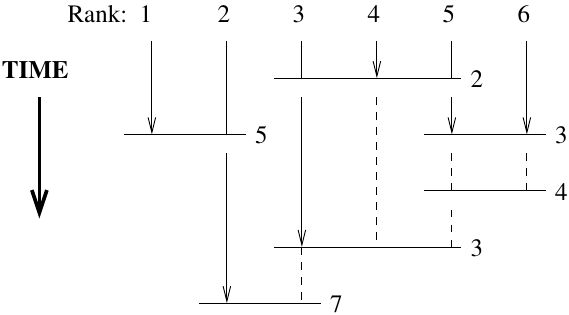}
\caption{Larger example of CC algorithm (updating the target sequence numbers)}
\label{fig:seq-num-algo-full}
\end{subfigure}
\caption{\small The two figures are examples of
snapshots in time.
An arrowhead in the timeline of an MPI process in Figure~\ref{fig:seq-num-algo-simple}
indicates the current point in time at which the checkpoint request arrived.
A solid vertical line is in the past and a dashed vertical line is
in the future of the MPI process.  A dashed vertical line terminates
at the collective operation that is a target for the given process.
A horizontal line indicates a (blocking) collective operation (a node
when viewing this as a directed graph).
The number to the right of each collective operation is the sequence
number assigned for that ggid (for the set of ranks of the group of
that operation).}
\label{fig:seq-num-algo}
\end{figure}

Figure~\ref{fig:seq-num-algo-simple} shows an example of target numbers.
The groups \{1,2\}, \{2,3\}, \{3,4,5\} and \{5,6\} (denoting ranks as
determined in \texttt{MPI\_COMM\_WORLD}) have
local targets 5, 7, 2, and 3, respectively.  Each individual process may
have multiple targets.  For example, rank~3 has a target of~2 for
the group~\{3,4,5\}, and a target of~7 for the group~\{2,3\}.  Process~3
has reached the target for the group~\{3,4,5\}, but not yet the
target for the group~\{2,3\}. In this case, process~3, 4 and 5 are allowed to continue executing until they finish targets 7, 2, and 3, respectively.

Figure~\ref{fig:seq-num-algo-full} shows a more complex case.
All processes have reached their targets, except for process~3.
Process~3 previously had a target of 7 for the group \{2,3\}.
But process~3 encountered and just executed a new operation on group \{3,4,5\}.
Process~3 increments its local sequence number, for this group,
and determines that the local sequence number~3 is larger than
the previous shared target of~2 for that group.
(\texttt{SEQ[ggid])>TARGET[ggid]} for this group.)

Since the CC algorithm requires at all times that
\texttt{max(SEQ[ggid])==TARGET[ggid]} (for the maximum
overall MPI processes), the complete CC algorithm
adds a new step.  The newly incremented \texttt{SEQ[ggid]}
must be shared with all other participating processes.

So, process~3
sends a message to processes~4 and~5, with the new local \hbox{sequence}
number,~3, for the group.  
    Note that process~3 can locally
discover the peer processes for the group \{3,4,5\} using the
local MPI call, \texttt{MPI\_Group\_translate\_ranks}.

Note that since process~5 has a new target for the group \{3,4,5\},
it will eventually execute the operation on group \{5,6\} again, and send
messages updating the target of that group to~4.  This will force
process~5 to start executing again.

\vskip25pt
\subsubsection{Blocking collective calls and point-to-point calls}\hfill
\label{sec:CCBlockingP2P}

\begin{figure}[h!t]
\centering
\includegraphics[width=0.9\columnwidth]{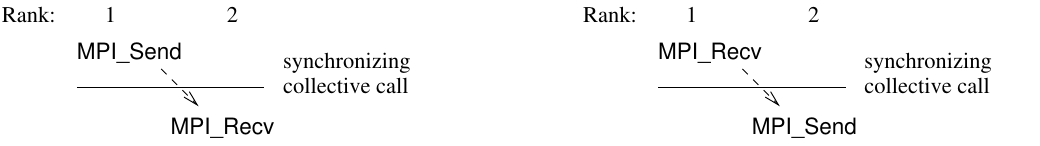}
\caption{The two cases above do \emph{not} occur in a correct MPI program.}
\label{fig:propertyB}
\end{figure}

 We recall again that blocking collective calls are assumed to be synchronizing (see
Section 3), and that ``an MPI collective procedure is synchronizing if it will only
return once all processes in the associated
group or groups of MPI processes have called the appropriate
matching MPI procedure.''~\cite[Section~2.4.2]{message2021mpi}.
Hence, a matched send-receive pair may not ``cross'' a blocking collective operation, as summarized in Figure~\ref{fig:propertyB}. 

\begin{comment}
The MPI standard forbids the above behavior in a correct MPI application.  Recall the following from the standard.
\begin{enumerate}
\item[(i)] The MPI standard~\cite[Section~6.1]{message2021mpi} states, ``Collective operations can (but are not required to) complete as soon as the caller's participation $\ldots$ is finished.''  Hence, an MPI implementation may choose to
complete a collective operation only when all participating processes have executed MPI\_Wait (or possibly
MPI\_Test).
\item[(ii)] The MPI standard~\cite[Section~3.4]{message2021mpi} further states, ``A program is `safe' if no message buffering [for a send-receive pair] is required for the program to complete.''
\end{enumerate}
\end{comment}

\subsection{The CC algorithm extended to non-blocking calls}
\label{sec:nonblocking}

    There are two issues to discuss:
    \begin{enumerate}
        \item How are \emph{SEQ[ggid]} and \emph{TARGET[ggid]} set when a collective
        operation may be non-blocking?
        \item What should be done when a ``safe'' point is reached, but no processes have tested the operation
        for completion?
    \end{enumerate}

\vskip25pt
\subsubsection{How are \emph{SEQ[ggid]} and \emph{TARGET[ggid]} set when a collective
        operation may be non-blocking?}\hfill
    
The pseudo-code of Section~\ref{sec:pseudo-code} updates \emph{SEQ[ggid]} and \emph{TARGET[ggid]}
at the time of a blocking collective communication call.  However, in the non-blocking
case, this call is now split between a call that initiates a non-blocking collective communication
call (e.g., \texttt{MPI\_Ibcast}) and a call that completes the non-blocking collective
call (e.g., \texttt{MPI\_Wait} or \texttt{MPI\_Test}).

According to the MPI standard, non-blocking collective communication operations are independent from
other MPI operations (both blocking and non-blocking) after they are initiated.
Therefore, at any point of time between the initiation and completion of a non-blocking collective operation, 
the operation may or may not be executing in the background. The CC algorithm assumes all initiated nonblocking collective operations immediately
start executing in the background.  Therefore, it increments the \texttt{SEQ[ggid]} during the
initiation phase. This approach guarantees that all possible messages in the network are received before the safe state for checkpointing.

    This choice to update during initiation and not later supports a common pattern. A process may initiate multiple non-blocking
collective communications at once, and wait for one or all processes to complete using functions like \texttt{MPI\_Waitany} and \texttt{MPI\_Waitall}.  (See~\cite[Example~6.35]{message2021mpi}.)

\vskip25pt
\subsubsection{What should be done if a ``safe'' point is reached,
but some processes haven't tested the nonblocking collective operation for completion?}\hfill

Since the CC algorithm increments the \emph{SEQ[ggid]} during the initiation of non-blocking communications,
it's possible that some of the communications haven't finished the communication when all processes reached all targets.

At a safe state, all processes of an incomplete non-blocking communication must have initiated the communication because of the invariant of a safe state.
Therefore, the communication will eventually complete if all processes start waiting for completion using functions like \texttt{MPI\_Test} and \texttt{MPI\_Wait}.
The CC algorithm keeps a list of \texttt{MPI\_Request} objects for incomplete non-band doelocking communications. When a safe state is reached, the CC algorithm will keep
calling \texttt{MPI\_Test} on each incomplete \texttt{MPI\_Request} until all communication have been completed.

\section{Experiments}
\label{sec:experiments}

All experiments were conducted on the Perlmutter Supercomputer at the National Energy Research Scientific
Computing Center (NERSC).  Perlmutter is the \#14 supercomputer on the Top-500 list
as of June, 2024~\cite{top5002024jun}.
Perlmutter has 3,072 CPU nodes and 1,792 GPU-accelerated nodes. Each CPU node has two AMD EPYC 7763 processors per node, for a total of 128~physical cores and 512~GB of RAM. The network uses HPE Cray's Slingshot~11 interconnect.
The Cray MPICH version is 8.1.25. Cray mpicc is based on gcc-11.2.
The Linux operating system is SUSE Linux Enterprise Server 15 SP4 (Release~15.4),
with Linux kernel~5.14.

Experiments are executed both on the older MANA using the 2PC subsystem for collective communication and also replacing this by the CC subsystem of this work.  (Note that MANA/2PC does not support non-blocking collective communication calls, and so that experimental comparison is not possible.  Because Perlmutter uses a modern network interconnect (Slingshot-11), it is not possible to compare MANA with the older transparent checkpointing packages based on MVAPICH~\cite{gao2006application} (2006) and Open~MPI~\cite{hursey2009interconnect} (2009).  Those earlier efforts supported only Ethernet and OFED InfiniBand.  And even the underlying BLCR on which they depend was last updated in January, 2013~\cite{blcrDownloads}.

The experiments are organized as follows.
Section~\ref {sec:micro-benchmarks} presents the OSU Micro-Benchmarks.  These micro-benchmarks represent an upper limit of the rate at which an application will make collective communication calls.  Those experiments show that at this extreme upper limit, the CC algorithm still performs reasonably, with a typical runtime overhead under 1.3\%, as opposed to the 2PC algorithm, whose runtime overhead frequently rises even beyond 100\%.

Section~\ref{sec:coll-call-per-sec} then discusses the rate of collective communication calls per second across all of the categories of applications analyzed here.

Section~\ref{sec:real-world} then analyzes the runtime overhead of five real-world applications.  As will be seen, it is at the higher rates of calls per second that 2PC has excessive runtime overhead, and the newer CC algorithm is required.

Finally, in Section~\ref{sec:vasp}, VASP~6 is analyzed in detail, since it showcases the performance of the CC algorithm for high rates of collective calls per second (2,489 calls per second).  VASP~6 is analyzed across a range of one to four computer nodes, as well as showing that checkpoint and restart times remain reasonable.

\subsection{Micro-benchmarks}
\label{sec:micro-benchmarks}

The OSU Micro-Benchmarks 7.0~\cite{network2022osu} were used to show the runtime overhead of the new CC algorithm compared to MANA's original two-phase-commit (2PC) algorithm.
Eight micro-benchmarks were chosen for different common patterns of blocking and non-blocking collective communications. Each experiment was repeated 5 times.

\begin{figure*}[h!t]
\centering
\begin{subfigure}{\textwidth}
   \centering
   \includegraphics[width=\textwidth]{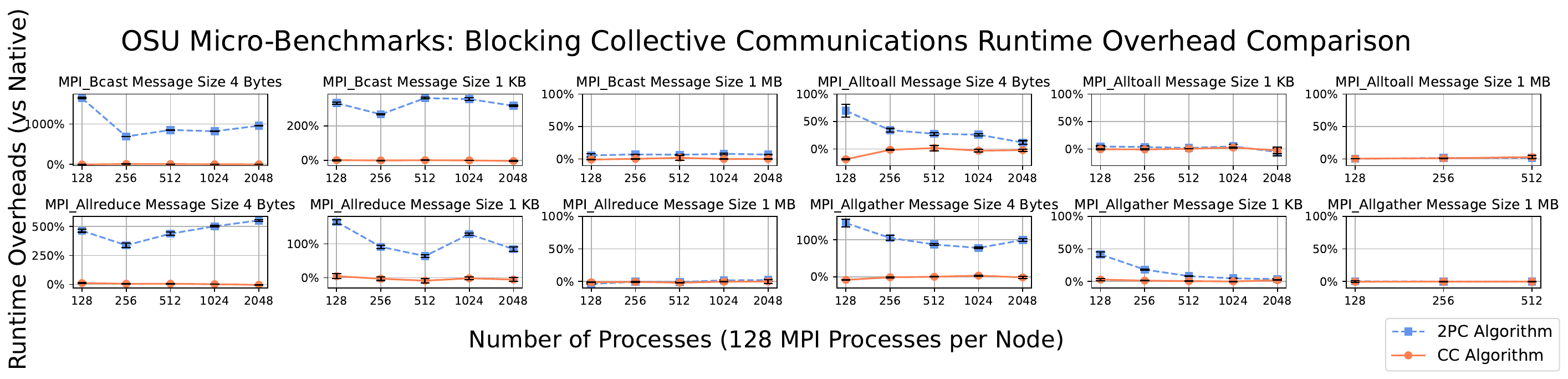}
  \caption{Blocking collective communications}
\end{subfigure}
\begin{subfigure}{\textwidth}
   \centering
   \includegraphics[width=\textwidth]{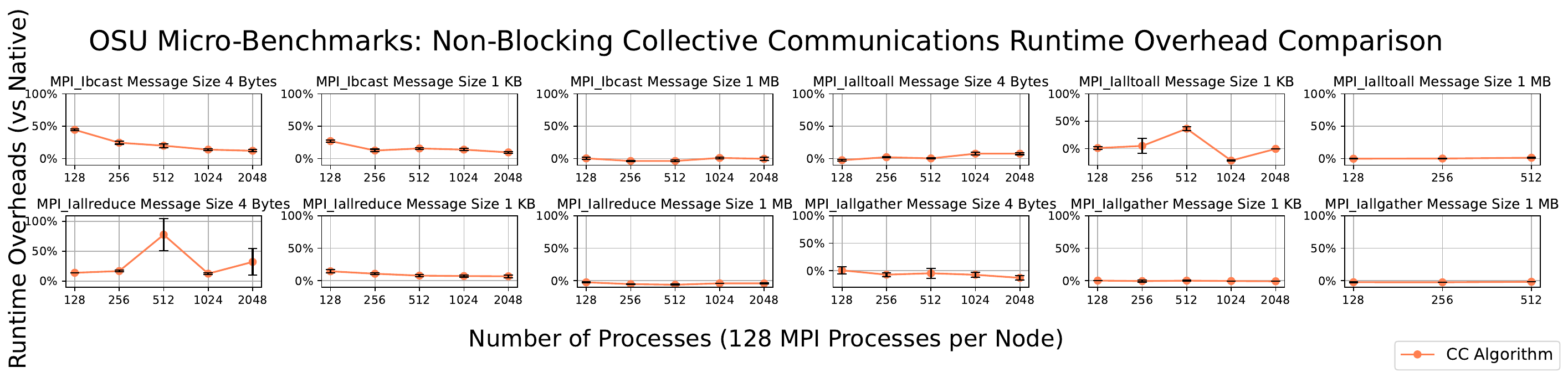}
  \caption{Non-Blocking collective communications}
\end{subfigure}
\caption{Runtime overhead on Micro-Benchmarks for CC and 2PC.  Note that 2PC is not shown for non-blocking functions since 2PC does not support such calls.}
\label{fig:micro_benchmark}
\end{figure*}

Figure~\ref{fig:micro_benchmark} shows the runtime overhead of both MANA's original two-phase-commit (2PC) and the CC algorithm of this work, with different message sizes (4~Bytes, 1~KB, and 1~MB). We scaled most micro-benchmarks up to 2048 processes over 16 nodes to test the scalability of the CC algorithm. The 2PC algorithm in the original MANA paper does not support non-blocking collective communication. Therefore, overhead is shown only for the CC algorithm, but not for 2PC.

MPI\_Alltoall/Ialltoall and MPI\_Allgather/Iallgather in the OSU Micro-Benchmarks do not support a message size of 1~MB over 1024 and 2048 processes, due to the default maximum memory limit. Hence, results are shown only up to 512 MPI processes for these cases.

\vskip25pt
\subsubsection{Blocking Collective Communication}\hfill
\label{sec:test_blocking}

The CC algorithm shows lower runtime overhead than 2PC in micro-benchmarks of blocking collective communications. In addition, the CC algorithm's runtime overhead remains consistently low as the message size or the number of nodes increases, whereas the runtime overhead of the 2PC algorithm varies depending on the message size and number of processes.

The 2PC algorithm inserts barriers that require extra communication and synchronization among processes. The additional communication increases the total latency of collective operations. Depending on the type of collective communication, the additional synchronization may have a different effect on performance. For example, MPI\_Bcast becomes slower because senders have to wait for all receivers to receive the message. But for functions like MPI\_Alltoall, the effect is minimal because the collective operation naturally requires synchronization among participating processors. 

In cases of large message size (1~MB), both algorithms perform identically to the native application. The cost of transferring messages is so large that the extra overhead introduced by each algorithm is insignificant.

\vskip25pt
\subsubsection{Non-blocking Collective Communication}\hfill
\label{sec:test_nonblocking}

The 2PC algorithm does not support non-blocking collective communications. Therefore, this section discusses runtime overhead for the CC algorithm only.

Note that for small messages, the runtime overhead for non-blocking communications is higher than for blocking counterparts and less stable. This is because the communication is divided into two phases: initiation and completion. CC has wrappers for the two phases, which contribute to the runtime. Therefore, the constant runtime overhead becomes larger than for the single wrapper of a blocking call. Nevertheless, the runtime overhead quickly decreases as the message size and number of nodes increase.

\begin{figure*}[h!bt]
\centering
\centering
\includegraphics[width=\textwidth]{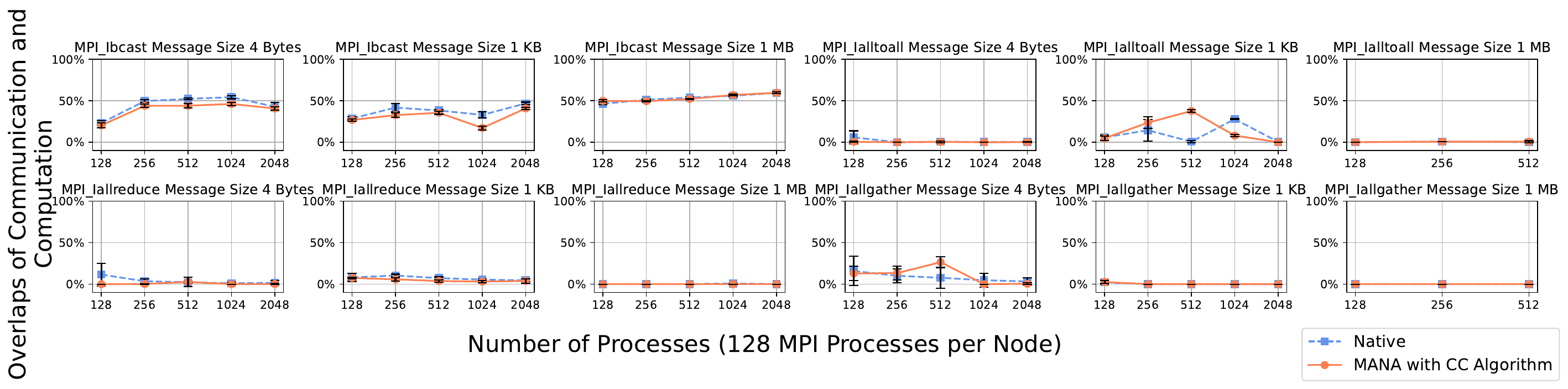}
\caption{Overlap of communication and computation when using non-blocking collective communications.}
\label{fig:overlaps}
\end{figure*}

Figure~\ref{fig:overlaps} shows percentages of overlap between communication and computation for non-blocking collective communication, as reported by the OSU Micro-Benchmarks. This overlap can improve the overall performance. The CC algorithm has a comparable amount of overlap compared to the native MPI implementation. Hence, the runtime overhead of CC in real-world programs is expected to be small, as seen in Sections~\ref{sec:real-world} and~\ref{sec:vasp}.

\subsection{Rates of Collective Communication Calls per Second}
\label{sec:coll-call-per-sec}

Table~\ref{tbl:inputs} shows the rate both of collective and point-to-point communication calls. Each experiment was run across four computer nodes, for a total of 512~processes.  The number of collective and point-to-point communication calls per second is computed as the average number of calls per second over all MPI processes. The OSU Micro-Benchmarks in Table~\ref{tbl:inputs} is a reference to the number of collective communications calls per second.

\begin{table*}[!ht]
\centering
\begin{tabular}{ |l|c|c|r|r|}
\hline
\textbf{Application}  & \textbf{Processes} & \textbf{Input} &\textbf{coll\hbox{.} comm\hbox{.}} & \textbf{point-to-point} \\
                      &                    &                &\textbf{calls/sec\hbox{.}}                      & \textbf{calls/sec\hbox{.}} \\
\hline
OSU MicroBench & 512 & MPI\_Bcast (msg: 4 bytes) & 255,754.5 & NA \\ \hline
VASP~6 & 512 & PdO4 & 2,489.2 & 2,568.9 \\ \hline
Poisson Solver & 512 & rel error = 0.01 & 21.3 & NA \\ \hline
CoMD  & 512 & Cu\_u6.eam & 7.8 & 414.2 \\ \hline
LAMMPS & 512 & Scaled LJ Liquid & 6.3 & 1,707.5 \\ \hline
SW4 & 512 & LOH.1-h50.in & 0.6 & 157.9 \\ \hline
\end{tabular}
\caption{\label{tbl:inputs} Input for each application, ordered by collective communication calls per second. All tests were conducted over 4 nodes on Perlmutter.} 
\end{table*}

It is clear that the runtime overhead of an MPI application depends critically on the number of collective communication calls per second.
Section~\ref{sec:real-world} will analyze the runtime overhead of five real-world applications.  The five applications can be categorized according to the rate of collective calls per second.  There are three categories: (i)~a low rate (less than 10~calls per second: SW4, LAMMPS, CoMD); (ii)~a medium rate (tens of calls per second: Poisson Solver); and (iii)~a very high rate (hundreds or thousands of calls per second: VASP~6).  The OSU Micro-Benchmark (hundreds of thousands of calls per second) is included as an upper limit.

As will be seen in the next subsection, for low rates of collective calls per second, CC and 2PC both have very low runtime overhead.  For a medium rate, unfortunately, the 2PC algorithm cannot support Poisson Solver, because this code uses non-blocking collective calls.  While CC still performs well, no overhead for 2PC could be exhibited.
One of the novelties of the CC algorithm is that it extends to support of non-blocking collective calls, as seen in Section~\ref{sec:nonblocking}.
For very high rates of calls per second, the CC algorithm widely outperforms the 2PC algorithm.  Finally, as was seen in the previous subsection, in the upper limit of the OSU Micro-Benchmarks, CC still performs well, while 2PC can exhibit more than 100\% runtime overhead.

\subsection{Real-world Applications}
\label{sec:real-world}

\begin{figure}[h!t]
\centering
\includegraphics[width=0.8\columnwidth]{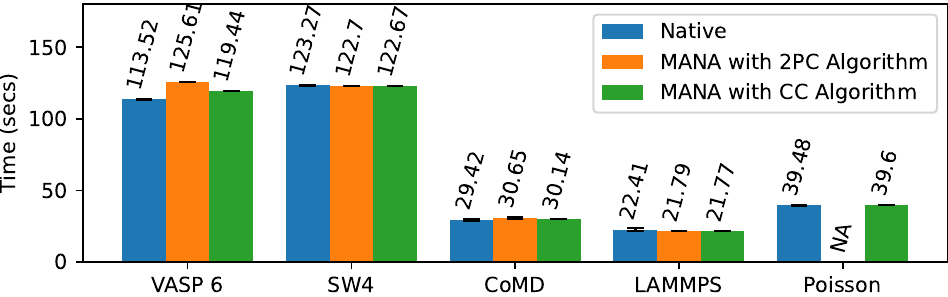}
\caption{Real-world applications runtime performance for 512 processes over 4 nodes.\\
(NOTE: Poisson uses non-blocking collective calls, supported by the newer CC, but not by 2PC.)}
\label{fig:real_world_overhead}
\end{figure}

We chose five real-world applications to show the CC algorithm's performance. 
Figure~\ref{fig:real_world_overhead} shows five real-world applications' runtime performance and standard deviation.  All experiments use 512 processes over 4~nodes on Perlmutter. Each test is repeated 5~times.

These applications have different communication patterns. Among the five applications, VASP~6 uses both collective communication and point-to-point communication most intensively (see Table~\ref{tbl:inputs}). Therefore, of all the real-world applications, VASP~6 places the greatest stress on runtime overhead. Nevertheless, even in this extreme case, CC algorithm achieves a runtime overhead of only~5.2\%, while the earlier 2PC algorithm has a runtime overhead of 10.6\%.

The Poisson Solver~\cite{hoefler2007optimizing} uses non-blocking collective communications only. Therefore 2PC is not applicable. The runtime overhead of the CC algorithm is less than 1\%.

In contrast,  CoMD~\cite{papa2001constrained}, LAMMPS~\cite{thompson2022lammps}, and SW4~\cite{sjogreen2012sw4} don't use collective communications frequently enough so that the runtime overhead of both CC and 2PC algotithms are negligible. 

The average runtime of SW4 and LAMMPS shown in the graphs indicates both CC and 2PC algorithms run slightly faster than the native applications (less than half a second), but within the standard deviation.

\subsection{Scalability of Real-world Application: VASP}
\label{sec:vasp}

\begin{figure}[htb!]
\centering
\includegraphics[width=0.8\columnwidth]{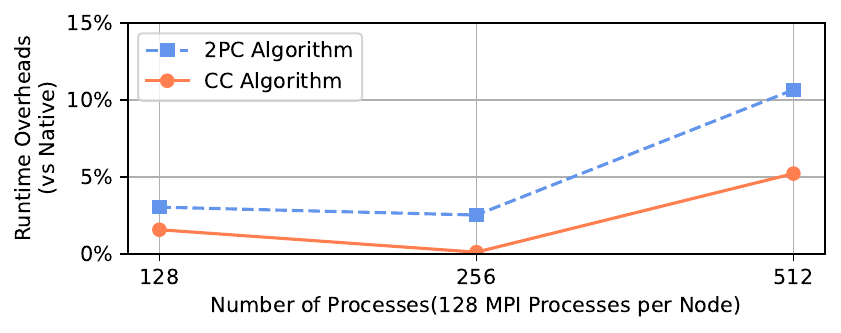}
\caption{VASP 6 Runtime Overhead: 2PC vs\hbox{.} CC}
\label{fig:vasp_overhead}
\end{figure}

\begin{figure}
\centering
\includegraphics[width=\columnwidth]{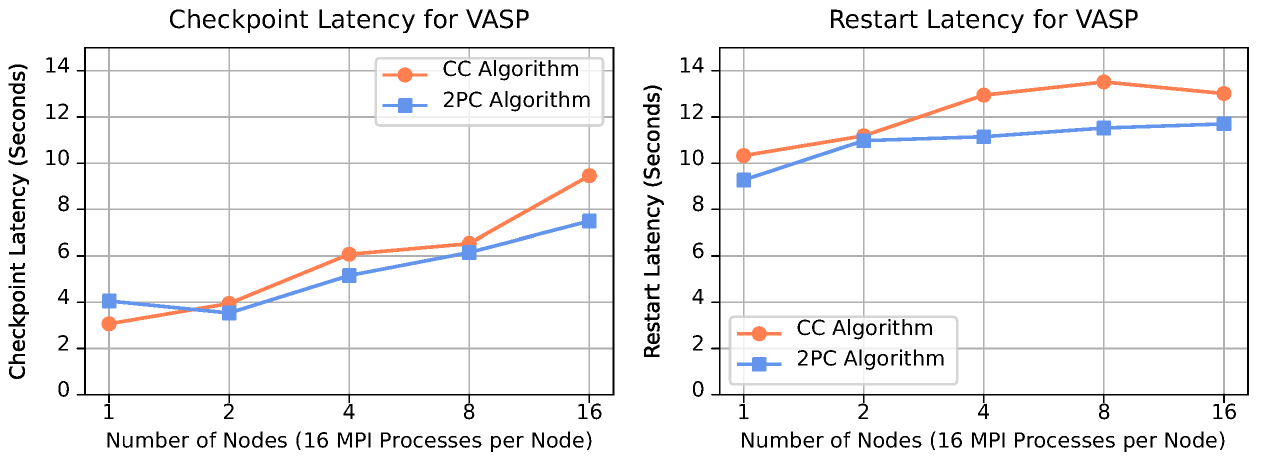}
\caption{VASP 6 Checkpoint and Restart Time: 2PC vs\hbox{.} CC}
\label{fig:vasp_ckpt_restart}
\end{figure}

We show the CC algorithm's scalability beyond micro-benchmarks in the case of VASP~6.
Figure~\ref{fig:vasp_overhead} shows the runtime overhead in each case. The result shows that the CC algorithm scales better than the earlier 2PC algorithm of the original MANA~\cite{garg2019mana}, ranging from a runtime overhead of 2\% (128 processes) to 5.2\% (512 processes). 

Both algorithms show a smaller runtime overhead with 256 processes than the result with 128. This is because the 128-process case uses a single node, and the 256-process case uses two nodes. When communicating between two physical nodes, the base cost of communication increases, and so the relative runtime overhead for two nodes is smaller than in the case of a single node.

As discussed in the introduction, VASP depends heavily on efficient FFTs, which requires low-latency networks.  When VASP runs on more nodes (e.g., above 4~nodes),
the efficiency of many workloads degrade. Therefore, we didn't test beyond 4 nodes since the performance results will no longer reflect the real-world usage of VASP.

Next, Figure~\ref{fig:vasp_ckpt_restart} shows that checkpoint and restart times for MANA/CC remain reasonable. This experiment was on a Lustre distributed file system. Each VASP processes uses about 700~MB memory.  The checkpoint command is issued at random times, and checkpoint-restart times are averaged over 5~runs.
In order to get realistic numbers, checkpoint-restart times are shown over a range from 1~node to 16~nodes.
Checkpoint and restart times are very close between the 2PC and CC
algorithms. For both 2PC and CC algorithms checkpoint and restart are slower when running on more nodes because there is more data in the memory that needs to be saved and restored.

Each checkpoint 
image file is 398~MB. The checkpoint image file is smaller than the memory usage because the lower half, which contains the MPI library and network drivers, is not saved in the image. 

The time to checkpoint depends strongly
on the bandwidth to stable storage. While the current times are
based on checkpointing to back-end disk-based Lustre distributed file system, newer
architectures are expected to show still better times, based on
the use of SSDs for intermediate storage.

\section{Related Work}
\label{sec:relatedWork}

A rich set of library-based packages for checkpointing MPI exists:
SCR~\cite{moody2010design} (2010), FTI~\cite{bautista2011fti} (2011), ULFM~\cite{bland2013post,losada2020fault} (2014), and Reinit~\cite{laguna2016evaluating} (2016, a simpler interface inspired by ULFM), and VeloC~\cite{nicolae2019veloc} (2019).

Transparent checkpointing for MPI also has a long history.
It was demonstrated in MPICH-V~\cite{bouteiller2006mpich}.  Soon after, BLCR~\cite{hargrove2006berkeley} was created to provide transparent checkpointing for a tree of processes on a single computer node.  BLCR was then leveraged to support transparent checkpointing of MPI.  The strategy was for an individual MPI implementation to: (a)~disconnect the network connection; (b)~use BLCR to transparently checkpoint each individual node; and (c)~finally to reconnect the network (or connect it for the first time, if restarting from checkpoint images).  This was done for InfiniBand by MVAPICH~\cite{gao2006application}, for Open~MPI~\cite{hursey2007design,hursey2009interconnect}, and for DMTCP~\cite{cao2014transparent}.  Unfortunately, the underlying BLCR software, itself, was last updated only in January, 2013~\cite{blcrDownloads}, and with the newer networks, transparent checkpointing is no longer supported.

MANA~\cite{garg2019mana} was then developed in 2019, using both split processes and the two-phase-commit algorithm for collective calls.  The current work shows how to replace the two-phase-commit algorithm with more efficient collective clocks.  Earlier developments of MANA were concerned with a production-quality version~\cite{chouhan2021improving,xu2021mana,xu2023implementation}.

The CC algorithm can be viewed as a \emph{consistent snapshot algorithm} for MPI collective operations.  Consistent snapshots to support point-to-point operations in the case of distributed algorithms include:  the original Chandy-Lamport algorithm~\cite{chandy1985distributed} and  Baldoni \hbox{et al.}~\cite{baldoni2001rollback}.

By analogy, the original Chandy-Lamport algorithm~\cite{chandy1985distributed} can be considered as a consistent snapshot algorithm for MPI point-to-point operations, although it does not apply to collective operations.
Indeed, Baldoni \hbox{et al.}~\cite{baldoni2001rollback} used similar ideas to demonstrate a rollback algorithm for transparent checkpointing of a distributed system with local checkpoints. Like Chandy-Lamport, their algorithm only supports point-to-point operations, and do not apply to collective operations.
Clocks and logical clocks~\cite{lamport1978time} themselves have a long history.  Two good introductions to the subject are~\cite{baldoni2002fundamentals} and~\cite{raynal1992logical}.

Collective communication has also been investigated within the domain of distributed systems.  Within distributed systems, a central concern is to make the implementation of collective communication fault-tolerant.  This is in comparison with MPI applications, where it is assumed that the underlying MPI library is responsible for fault tolerance.

Some example studies of fault tolerance for collective communication in distributed systems include Hoplite~\cite{zhuang2021hoplite}, and scalable distributed collectives for Asynchronous Many-Task (AMT) models~\cite{whitlock2018scalable}, 

\section{Conclusion}
\label{sec:conclusion}
The new CC algorithm reduces the runtime overhead for VASP from up to 10\% in MANA's old two-phase-commit (2PC) algorithm to typically 5\% or less in the new algorithm.  Similarly, the OSU Micro-Benchmark for blocking collective calls shows a drastic improvement in this stress test (with up to 2048 processes):  from above 100\% runtime overhead with the old 2PC algorithm to nearly 0\% with the new collective-clock (CC) algorithm.  For non-blocking collective calls, the older 2PC algorithm does not support that case.  But the CC algorithm executes at a runtime overhead typically between 0\% and 10\%, for up to 2048 processes.
An atypical worst case occurs for \texttt{MPI\_Ibcast}, where CC can show runtime overheads of up to 50\%.

\section*{Acknowledgment}
We wish to thank Zhengji Zhao for her encouragement and helpful insights.  We also wish to thank both NERSC at Lawrence Berkeley National Laboratory and MemVerge, Inc. for the use of their facilities.  Finally, we wish to thank Kapil Arya for his enhancement of MANA by correctly tagging the memory regions of the upper and lower halves.  This enhancement allowed us to remove the MPICH\_SMP\_SINGLE\_COPY\_OFF environment variable, thus improving the runtime overhead in all experiments, both for the original MANA code and for the newer algorithm.  And we thank Twinkle Jain for valuable discussions.

% \bibliographystyle{ieeetr}
% \bibliography{ref}

\end{document}